\documentclass[12pt]{iopart}

\usepackage[lmargin=2.5cm,rmargin=2.5cm,tmargin=3cm,bmargin=2cm]{geometry}
\usepackage{amssymb}
\usepackage{graphicx}
\usepackage{graphics}
\usepackage{appendix}
\usepackage{multirow}
\usepackage{hhline}
\usepackage[normalem]{ulem}
\usepackage[usenames, dvipsnames]{color}
\usepackage{relsize}
\usepackage{color}
\usepackage[font=small]{caption}
\usepackage{subcaption}
\usepackage{xspace}
\usepackage{cite}
\usepackage{hyperref}
\hypersetup{
    colorlinks=true,
    linkcolor=blue,
    filecolor=blue,      
    urlcolor=blue,
    citecolor=blue
}

\usepackage[capitalize]{cleveref}
\newcommand{\halfwidth}{0.49\textwidth}

\newcommand{\RE}{RE\xspace}
\newcommand{\REs}{REs\xspace}
\newcommand{\TQ}{TQ\xspace}
\newcommand{\CQ}{CQ\xspace}

\newcommand{\SPARC}{SPARC\xspace}
\newcommand{\REMC}{REMC\xspace}
\newcommand{\DREAM}{DREAM\xspace}
\newcommand{\PRD}{PRD\xspace}

\newcommand{\DIIID}{DIII-D\xspace}
\newcommand{\NIMROD}{NIMROD\xspace}
\newcommand{\ASCOT}{ASCOT5\xspace}

\newcommand{\MHD}{MHD\xspace}
\newcommand{\COMSOL}{COMSOL\xspace}

\newcommand{\VV}{VV\xspace}

\newcommand{\Poincare}{Poincar\'{e}\xspace}
\newcommand{\ohmic}{Ohmic\xspace}
\newcommand{\ala}{\`{a} la\xspace}

\newcommand{\pmc}{p/m_e c}
\newcommand{\A}{A}
\newcommand{\D}{D}

\newcommand{\n}{n}

\renewcommand{\t}{t}
\newcommand{\Ip}{I_\mathrm{p}}
\newcommand{\Bo}{B_0}
\newcommand{\Ro}{R_0}

\newcommand{\Wmag}{W_\mathrm{mag}}
\newcommand{\li}{\ell_\mathrm{i}}
\newcommand{\ppar}{p_\parallel}

\newcommand{\db}{\delta B / B}

\newcommand{\SI}[2]{#1\,\mathrm{#2}}

\newcommand{\psiN}{\psi_\mathrm{N}}
\newcommand{\q}{q}
\newcommand{\qo}{\q(0)}
\newcommand{\Teo}{T_{e0}}
\newcommand{\mysim}{{\sim}}
\newcommand{\mydash}{{-}}

\newcommand{\sub}[1]{\sout{#1}}
\renewcommand{\sub}[1]{\unskip}

\newcommand{\enadd}[1]{\textcolor{red}{#1}}
\renewcommand{\enadd}[1]{#1\unskip}
\newcommand{\ensub}[1]{\sout{#1}}
\renewcommand{\ensub}[1]{\unskip}

\newcommand{\ppcfadd}[1]{\textcolor{red}{#1}}
\renewcommand{\ppcfadd}[1]{#1\unskip}
\newcommand{\ppcfsub}[1]{\sout{#1}}
\renewcommand{\ppcfsub}[1]{\unskip}

\begin{document}
    
    \title[]{On the minimum transport required to passively suppress runaway electrons in \SPARC \ppcfadd{disruptions}}%{On the minimal transport required to passively suppress runaway electrons in \SPARC}

\newcommand{\iPSFC}{1\xspace}
\newcommand{\iChalmers}{2\xspace}
\newcommand{\iFiatLux}{3\xspace}
\newcommand{\iIPP}{4\xspace}
\newcommand{\iEPFL}{5\xspace}
\newcommand{\iColumbia}{6\xspace}

\newcommand{\PSFC}{$^\iPSFC$ Plasma Science and Fusion Center, Massachusetts Institute of Technology, Cambridge, MA, USA\xspace}
\newcommand{\Chalmers}{$^\iChalmers$ Department of Physics, Chalmers University of Technology, SE-41296 G\"{o}teborg, Sweden}
\newcommand{\FiatLux}{$^\iFiatLux$ Fiat Lux, San Diego, CA 92101, USA}
\newcommand{\IPP}{$^\iIPP$ Max Planck Institute for Plasmaphysics, 85748 Garching, Germany}
\newcommand{\Columbia}{$^\iColumbia$ Department of Applied Physics and Applied Mathematics, Columbia University, NY, USA}
\newcommand{\EPFL}{$^\iEPFL$ Ecole Polytechnique F\'{e}d\'{e}rale de Lausanne (EPFL), Swiss Plasma Center (SPC), CH-1015 Lausanne, Switzerland} %

\author{R.A.~Tinguely\iPSFC\footnote{Author to whom correspondence should be addressed: rating@mit.edu}, 
    I~Pusztai$^\iChalmers$,
    VA~Izzo$^\iFiatLux$,
    K~S\"{a}rkim\"{a}ki$^\iIPP$,
    T~F\"{u}l\"{o}p$^\iChalmers$,
    DT~Garnier$^\iPSFC$,
    RS~Granetz$^\iPSFC$,
    M~Hoppe$^{\iEPFL}$,
    C~Paz-Soldan$^\iColumbia$,
    A~Sundstr\"{o}m$^\iChalmers$,
    and
    R~Sweeney$^\iPSFC$}
    
    \address{\PSFC \\
             \Chalmers \\
             \FiatLux \\
             \IPP \\
             \EPFL \\
             \Columbia}
    \begin{abstract}
    In [V.A. Izzo \emph{et al} 2022 \emph{Nucl. Fusion} \textbf{62} 096029], state-of-the-art modeling of thermal and current quench (\CQ) \MHD coupled with a self-consistent evolution of runaway electron (\RE) generation and transport showed that a non-axisymmetric ($n = 1$) in-vessel coil could passively prevent \RE beam formation during disruptions in \SPARC, a compact high-field tokamak projected to achieve a fusion gain $Q > 2$ in DT plasmas. However, such suppression requires finite transport of \REs within magnetic islands and re-healed flux surfaces; conservatively assuming \emph{zero} transport in these regions leads to an upper bound of \RE current $\mysim\SI{1}{MA}$ \ppcfadd{compared to $\mysim\SI{8.7}{MA}$ of pre-disruption plasma current}. Further investigation finds that core-localized electrons, within $r/a < 0.3$ and with kinetic energies $\mysim\SI{0.2\mydash15}{MeV}$, contribute most to the \RE plateau formation. Yet only a relatively small amount of transport, i.e. a diffusion coefficient $\mysim \SI{18}{m^2/s}$, is needed in the core to fully mitigate these \REs. Properly accounting for (i)~the \CQ electric field's effect on \RE transport in islands and (ii)~the contribution of significant \RE currents to disruption \MHD may help achieve this.   
\end{abstract}


% Varenna 2022 abstract
% In [1], state-of-the-art modeling of disruption MHD coupled with a self-consistent evolution of runaway electron (RE) generation and transport showed that a non-axisymmetric (n = 1) in-vessel coil can passively prevent RE beam formation during disruptions in SPARC, a compact high-field tokamak projected to achieve a fusion gain Q > 2 in DT plasmas. However, this analysis assumed a tightly fitting ideal wall, unconstrained coil current (up to 590 kA), and an artificially triggered thermal quench (TQ) with no MHD-driven losses. In this talk, we present the results of extended modeling [2] which captures additional physics compared to the reference case: A longer current quench (CQ) is achieved with a more realistic wall location or higher CQ temperature, each leading to higher RE loss than avalanche growth rates. Exploring the effect of the coil current, we find that a limit of 250 kA can allow a RE plateau to form with current ~2 MA depending on the choice of other model parameters. Finally, the TQ and coil work together to enhance RE loss early on, but the resulting late-CQ safety factor (q) profile is non-resonant with the coil, leading to a RE plateau ~1.5 MA. Importantly, the contribution of RE current to the safety factor is not yet included, but this could reduce q, regain resonance, and induce further MHD and transport.
    
    %\pagebreak
    %\ioptwocol
    
    \section{Introduction}\label{sec:intro}

    In \cite{Boozer2011,Smith2013}, a novel method was proposed for \emph{passive} mitigation of relativistic ``runaway electrons'' (\REs) generated during tokamak plasma disruptions: First, an in-vessel, non-axisymmetric coil would be passively energized through mutual coupling to the plasma current during the disruption's current quench (\CQ); then, the resulting magnetic field perturbation would enhance stochasticity and transport such that the \RE loss rate would dominate the growth rate, thus preventing \RE beam formation.
    
    In \cite{Sweeney2020}, such a ``Runaway Electron Mitigation Coil'' (\REMC) was proposed for the \SPARC tokamak \cite{Creely2020}, a high-field ($\Bo = \SI{12.2}{T}$), compact ($\Ro = \SI{1.85}{m}$, $a = \SI{0.57}{m}$) device currently under construction in Devens, Massachusetts, USA. The present \REMC design has a predominantly $\n=1$ structure and is located on the outboard wall; a similar coil is planned for the \DIIID tokamak, but on the inboard wall \cite{Weisberg2021}. Several aspects of the \SPARC ``Primary Reference Discharge'' (\PRD) make the \RE problem challenging: a large plasma current ($\Ip = \SI{8.7}{MA}$) can lead to dangerous exponential \RE growth; high core temperatures $\Teo \approx \SI{20}{keV}$ can cause enhanced primary and hot-tail generation; DT fuel provides a seed of non-thermal electrons through tritium beta decay; and high energy gammas from activated materials could accelerate electrons via Compton scattering.
    
    However, in \cite{Tinguely2021}, modeling of the \PRD's worst-case-scenario \CQ ($\mysim\SI{3}{ms}$) showed complete prevention of \RE beam formation with the \REMC~-- and $\mysim\SI{5\mydash6}{MA}$ of \RE current without it. The modeling workflow included four steps: First, the mutual couplings of all toroidally conducting structures were simulated in \COMSOL \cite{comsol} to evaluate the \REMC's vacuum electromagnetic fields during the worst-case \CQ. Second, these magnetic fields were applied at the boundary of a nonlinear, 3D \NIMROD \cite{sovinec:2004} simulation to assess the plasma response and total fields. Third, the stochastic magnetic fields were input into the orbit-following code \ASCOT to calculate the advective and diffusive transport \cite{hirvijoki2014ascot} of energetic electrons. Finally, these transport coefficients -- $\A,\D$ as functions of energy, pitch, and radius -- were supplied to the hybrid fluid-kinetic code \DREAM \cite{Hoppe_DREAM-2021} for self-consistent evolution of the \RE population. Importantly, in both \NIMROD and \DREAM, the \REMC vacuum fields and transport coefficients, respectively, were evolved as functions of $\Ip$ and not time explicitly. 
    
    More recently, in \cite{Izzo2022}, both the thermal quench (\TQ) and \CQ were modeled for the \SPARC \PRD and \REMC; the results of this study -- which bound the maximum expected \RE current -- are summarized in \cref{sec:Val}. \Cref{sec:Istvan} further explores these bounds in \RE phase space, as well as the minimum transport needed to fully prevent \RE beam formation. Finally, results and opportunities for future modeling are discussed in \cref{sec:discussion}.%, and a summary is provided in \cref{sec:summary}.
    
    %Finally, the implications for future modeling and \REMC operation are discussed in \cref{sec:discussion}.%, and a summary is provided in \cref{sec:summary}.
    \section{\REMC efficacy during the thermal and current quenches}\label{sec:Val}

\begin{figure}[h!]
        \centering
        \begin{subfigure}{\halfwidth}
            \includegraphics[width=\textwidth]{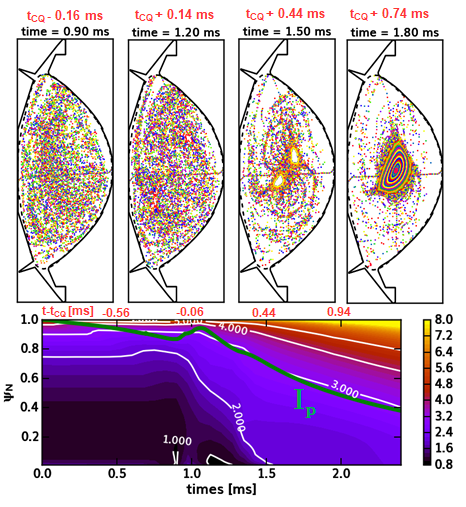}%{figs/SPARC_qprof+nimfl_g2.png}
            \caption{}%\NIMROD results.}
            \label{fig:NIMROD}
        \end{subfigure}
        \begin{subfigure}{\halfwidth}
            \includegraphics[width=\textwidth]{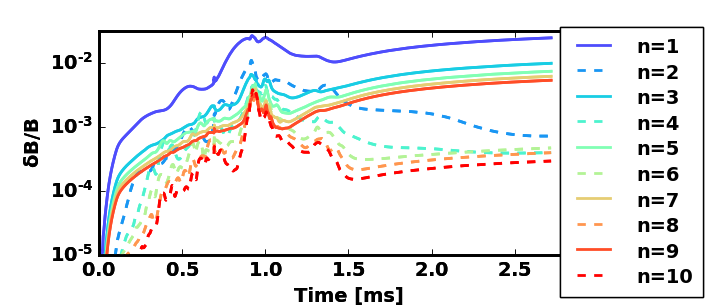}%{figs/dBoverB_TQ+CQ_PPCF.png}
            \caption{}%\DREAM results.}
            \label{fig:dB}
        \end{subfigure}
        %\begin{subfigure}{\halfwidth}
        %    \includegraphics[width=\textwidth]{figs/semilogIpPlotTQCQsim_SauterFull5.pdf}
        %    \caption{\DREAM results.}
        %    \label{fig:DREAM}
        %\end{subfigure}
        \caption{
            (a, upper)~Poincar\'{e} plots of (mostly) stochastic magnetic field lines from \NIMROD within the simulation boundary (dashed) and \SPARC first wall (solid). (a, lower)~The safety factor $\q$-profile evolution vs normalized poloidal flux ($\psiN$) and time, with the plasma current ($\Ip$) time-evolution overlaid.
            \ppcfadd{(b)~Amplitudes of $\n=1{-}10$ modes in units of $\db = \sqrt{\Wmag(\n)/\Wmag(\n=0)}$ with $\Wmag$ the magnetic energy Fourier component.}
            %(b)~Time-traces of \ohmic (solid), \RE (dot-dashed), and total (dashed) currents from \DREAM; thick/thin \RE currents indicate no/transport within re-healed flux surfaces. Times denoted above (b) correspond to the \NIMROD simulation. Transport coefficients are fixed in time after the vertical dashed line (b, upper), and surface re-healing begins at the vertical dotted line (b, lower). Subplots (a) and (b) are reproduced from Figures~5 and 6, respectively, in \cite{Izzo2022}.
            Subplot (a) is reproduced from Figure~5 in \cite{Izzo2022}.
        }
        \label{fig:Val}
    \end{figure}

\begin{figure}[h!]
        \centering
        \includegraphics[width=\halfwidth]{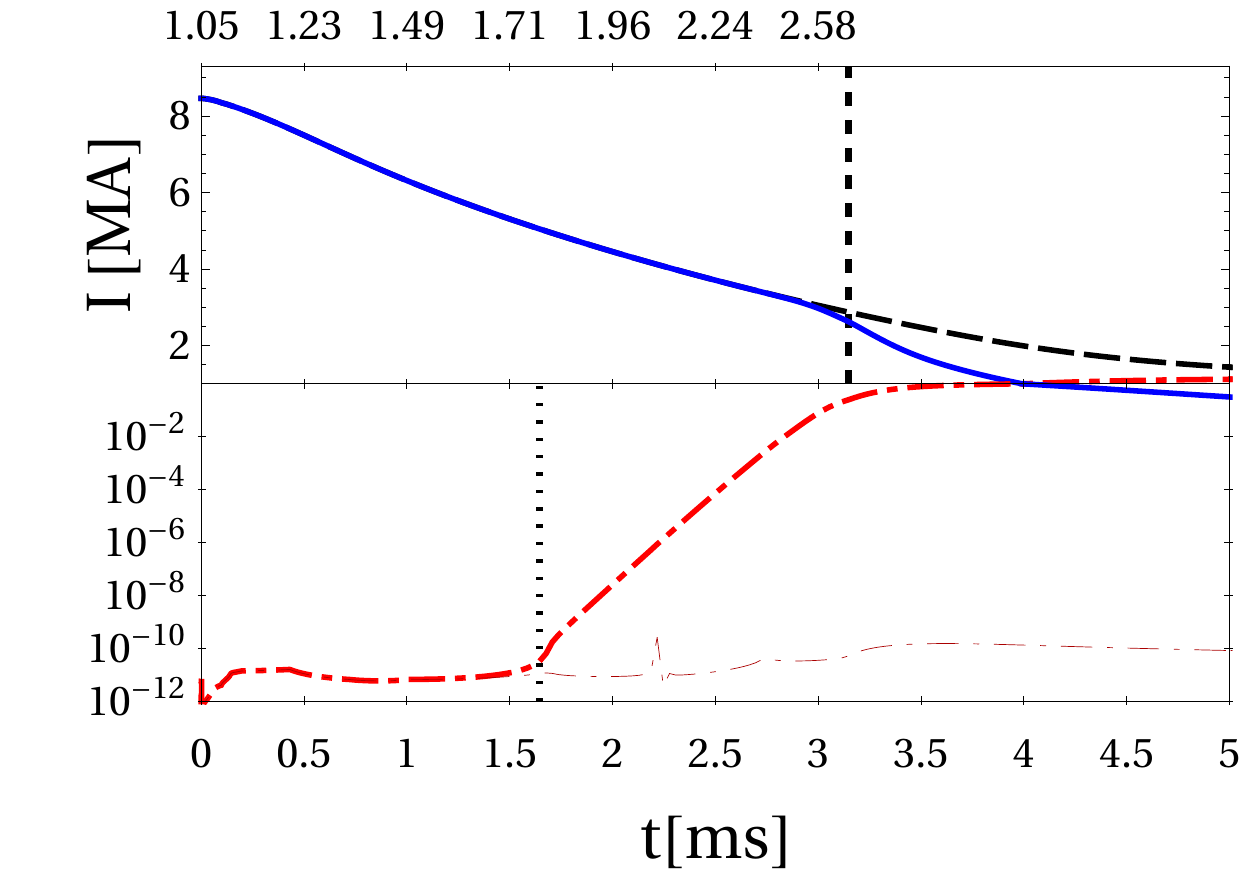}%{figs/semilogIpPlotTQCQsim_SauterFull5.pdf}
        \caption{Time-traces of \ohmic (solid), \RE (dot-dashed), and total (dashed) currents from \DREAM; thick/thin \RE currents indicate no/transport within re-healed flux surfaces. Times denoted above correspond to the \NIMROD simulation (see \cref{fig:Val}). Transport coefficients are fixed in time after the vertical dashed line (upper), and surface re-healing begins at the vertical dotted line (lower). Reproduced from Figure~6 in \cite{Izzo2022}.}
        \label{fig:DREAM}
    \end{figure}

    The same workflow presented in \cite{Tinguely2021} and summarized in \cref{sec:intro} was used in \cite{Izzo2022} to assess the \SPARC \REMC's efficacy for a full \PRD mitigated disruption, i.e. including both the \TQ and \CQ. Here, the \TQ \ppcfadd{($\mysim\SI{1}{ms}$ in duration)} was induced by neon radiation, as in a scenario where massive gas injection was employed. The main results are captured in \cref{fig:Val,fig:DREAM}. The pre-disruption safety factor ($\q$) profile is shown at $\t=0$ in \cref{fig:NIMROD} with $\qo \sim 1$ and $\q=2$ around a normalized poloidal flux value of $\psiN \approx 0.75$. During the \ensub{the} \TQ, i.e. the first $\mysim\SI{1}{ms}$ of the simulation, the plasma current $\Ip$ decreases slightly, with the $\Ip$-spike denoting the start of the \CQ. \ppcfadd{Around that time, the magnetic perturbation amplitudes $\db$ first peak (see \cref{fig:dB}), and strong nonlinear coupling among odd and even toroidal harmonics is observed.} \Poincare plots of magnetic field lines\ppcfadd{, in \cref{fig:NIMROD}, also} show high stochasticity during this period. 

    However, from $\t \approx \SI{1\mydash1.5}{ms}$, $\qo$ increases from 1 to 2, and beyond $\t > \SI{1.5}{ms}$, the \REMC is no longer resonant with the plasma core (refer to Figure~1 in \cite{Tinguely2021} for more details). \ppcfadd{Although the predominantly odd externally applied fields continue to grow as a the coil current continues to increase, these are now largely non-resonant fields that do not perturb the flux surfaces, and the nonlinearly excited resonant field components, both odd and even, decay away.} Thus, small islands start to reform, re-healing as closed flux surfaces by $\t \approx \SI{1.8}{ms}$. Note that the contribution from \REs to the \MHD are \emph{not} included in these \NIMROD simulations, \ensub{although} \enadd{but} the back-reaction is expected to be small for low \RE currents early on. This will be discussed further in \cref{sec:discussion}.
    
    \Cref{fig:DREAM} shows the self-consistent evolution of \ohmic and \RE currents from \DREAM, including the advective and diffusive transport calculated by \ASCOT in \DREAM's fluid transport model \cite{svensson_2021}. Note that the time bases of the \DREAM and \NIMROD simulations are not exactly the same; instead, the \DREAM simulation is initialized with profiles close to the time of \NIMROD's $\Ip$-spike. 
    \ppcfadd{
        Importantly, the \TQ in \DREAM is only modeled for the final $\mysim\SI{0.1}{ms}$ of \NIMROD's $\mysim\SI{1}{ms}$ \TQ because \DREAM requires a monotonic variation of the plasma current from which to map transport coefficients. These
    }%
    transport coefficients evolve with the plasma current until the final $\Ip$-value of the \NIMROD simulation; then, they are held constant in time (see the vertical dashed line in the upper part of \cref{fig:DREAM}).%
        \footnote{
            \ppcfadd{To enforce transport ambipolarity in \DREAM, any change of the electron density on a given flux surface due to transport is compensated by a change in the bulk electron density of similar size but opposite sign.}
        }

    Two scenarios for the \RE current are depicted in \cref{fig:DREAM}: In the first, the transport coefficients are applied as calculated throughout the entire plasma domain, i.e. even inside the re-healed flux surfaces, and the \RE current remains negligibly low ($\mysim\SI{1}{\mu A}$). However, in the second case, transport inside islands \enadd{and re-healed flux surfaces} is set to zero, which is perhaps overly conservative. (Explicitly, $A = D = 0$ wherever $\D<\SI{1000}{m^2/s}$.) The result is a \RE plateau with current $\mysim\SI{1}{MA}$. While this value is an improvement upon the $\mysim\SI{5\mydash6}{MA}$ of \RE current expected with \emph{no} \REMC \cite{Sweeney2020,Tinguely2021}, it is likely the pessimistic upper bound on the true value. 

    \ppcfadd{
        Here, it is important to note that the initial hot-tail seed can be sensitive to the \TQ cooling time prescribed in \DREAM. In the \CQ-only modeling effort of \cite{Tinguely2021}, a $\mysim\SI{0.5}{ms}$ \TQ from $\SI{20}{keV}$ to $\SI{4}{eV}$ led to a $\mysim\SI{50}{kA}$ seed which was then dissipated by the \REMC, with similar $\db \sim 10^{-2}$ as that in \cref{fig:dB}. Further reducing the \TQ time to $\SI{0.1}{ms}$ resulted in a much higher transient \RE beam current of $\mysim\SI{1}{MA}$, which was still suppressed by the \REMC. These results are consistent with all test-particle \REs losing confinement during the \TQ in \NIMROD for the scenario modeled in this paper \cite{Izzo2022}. The inclusion of \TQ transport in \DREAM, as a function of non-monotonic $\Ip$ variation, is left for future work.
    }
    \section{Investigating transport inside re-healed flux surfaces}\label{sec:Istvan}

\begin{figure}[h!]
        \centering
        \begin{subfigure}{\halfwidth}
            \includegraphics[width=\textwidth]{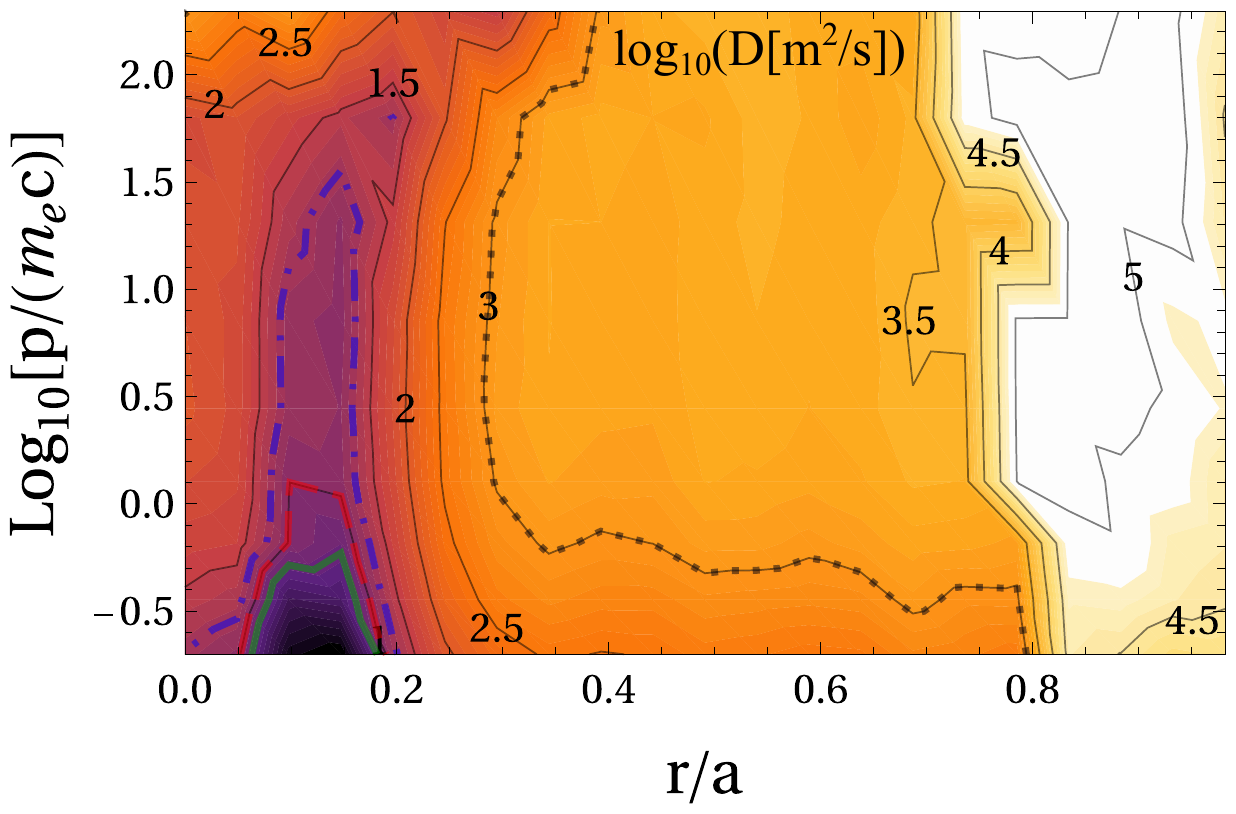}%{figs/FilteredOut2.pdf}
            \caption{\ASCOT results.}
            \label{fig:ASCOT}
        \end{subfigure}
        \begin{subfigure}{\halfwidth}
            \includegraphics[width=\textwidth]{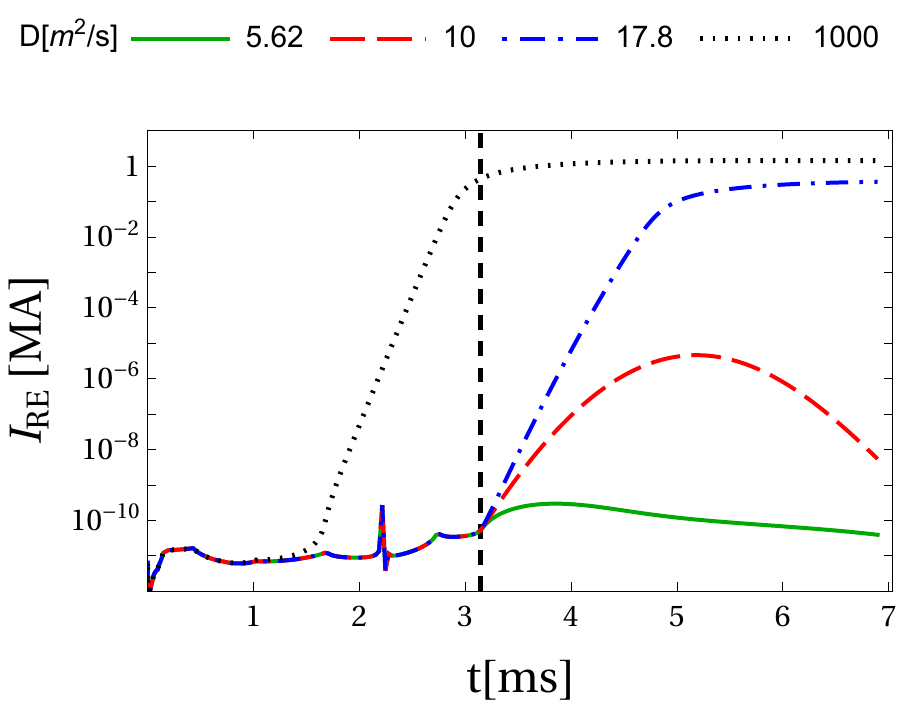}%{figs/IReLog.pdf}
            \caption{\DREAM results.}
            \label{fig:IrLog}
        \end{subfigure}
        \caption{(a)~Diffusion coefficients, $\log_{10}(\D\SI{}{[m^2/s]})$, from \ASCOT vs normalized minor radius ($r/a$) and electron momentum normalized to the rest mass ($\pmc$) at the time indicated by the vertical dashed lines in \cref{fig:DREAM}(upper) and subplot~(b). (b)~Time-traces of the \RE current from \DREAM when diffusion coefficients \emph{less} than the noted value are set to \emph{zero}, i.e. $\D=0$ \emph{within} the similarly styled contours in (a). Note the various linear/logarithmic scales. The legend for curves in subplot~(b) applies to contours in subplot~(a).} %(b)~Time-traces of the \RE current from \DREAM when \emph{lesser} diffusion coefficients are set to \emph{zero} \emph{within} the similarly styled contours in (a). Note the various linear/logarithmic scales. The legend for curves in subplot~(b) applies to contours in subplot~(a).}
    \end{figure}

    This section explores further the ten-order-of-magnitude difference in the predicted \RE current when transport is/not accounted for within \NIMROD's islands and re-healed flux surfaces. Radial profiles of the diffusion coefficient ($D$) are shown in \cref{fig:ASCOT}, also as a function of normalized electron momentum, at the last \NIMROD simulation time; the values shown are taken at a representative electron pitch $\ppar/p=0.8$. There are a few important notes here: (i)~the diffusion coefficients span five orders of magnitude from the plasma core to edge; (ii)~though not shown, the advection coefficients are of similar magnitude ($\A[{\rm m/s}] \sim \D[{\rm m^2/s}]$); and (iii)~both transport coefficients are relatively insensitive to the electron pitch in the relevant range $\ppar/p\in [0.8,\,1]$ (see Figure~3 in \cite{Tinguely2021}). 

    In \cref{fig:ASCOT}, general trends are seen of rapidly decreasing transport with decreasing radius and relative insensitivity to electron energy. However, there is a clear feature of ``very low'' transport ($\D < \SI{30}{m^2/s}$) for electrons localized in the core ($r/a \sim 0.05\mydash0.2$) and with energies ${<}\SI{50}{MeV}$ ($\pmc < 100$). %($\log_{10}(\pmc) \sim -0.75\mydash2$). 
    \Cref{fig:IrLog} shows the time-evolution of \RE current when the transport coefficients are zeroed in different regions of the phase space in \cref{fig:ASCOT}.%
        \footnote{Note that the diffusivity is used for discrimination of the phase space regions, while the advection coefficients (not shown here) are also filtered in the same regions.} 
    The ``base case'' is $\D=0$ wherever $\D < \SI{1000}{m^2/s}$, which effectively includes the entire core, $r/a < 0.3$, and leads to the previously seen $\mysim\SI{1}{MA}$ \RE beam. Yet reducing this threshold to $\D < \SI{10}{m^2/s}$ leads to negligible \RE current. Thus, it is primarily the electron population within $\D \sim \SI{10\mydash18}{m^2/s}$, i.e. localized in $r/a \sim 0.05\mydash0.2$ and with kinetic energies $\mysim\SI{0.2\mydash15}{MeV}$ ($\pmc \sim 1\mydash30$), which contributes most to \RE plateau formation.
    
\begin{figure}[h!]
        \centering
        \includegraphics[width=\halfwidth]{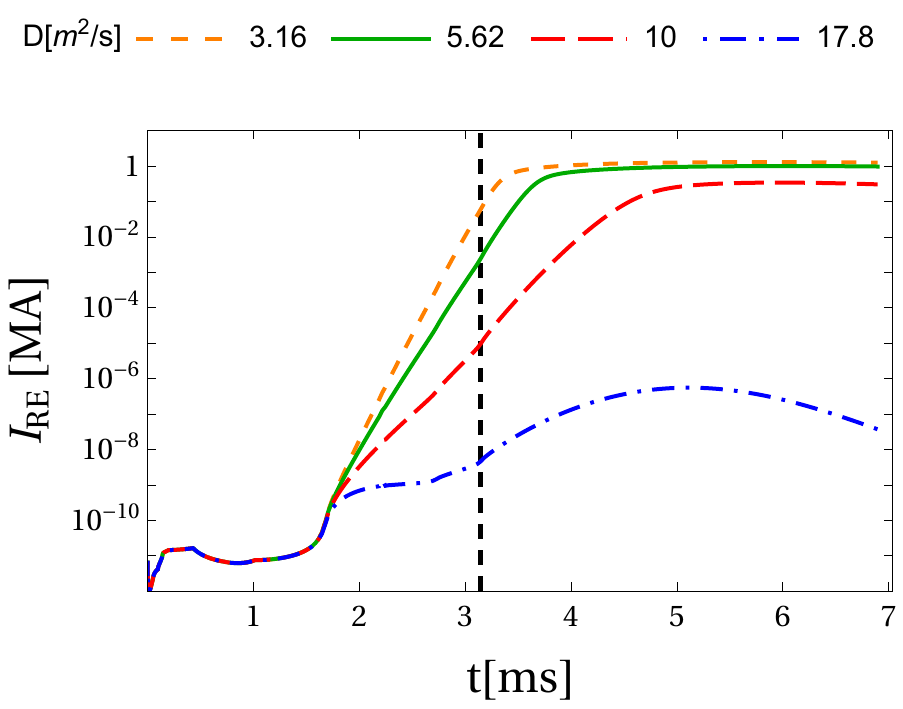}%{figs/IReLogInverse.pdf}
        \caption{Time-traces of the \RE current from \DREAM when the diffusion coefficient is set to the listed value in regions of phase space with $\D<\SI{1000}{m^2/s}$ in \cref{fig:ASCOT}. The time indicated by the vertical dashed line is the same as in \cref{fig:DREAM,fig:IrLog}.}
        \label{fig:IrLogInverse}
    \end{figure}

    This problem can be looked at from another angle: What is the minimum transport needed to fully suppress \RE plateau formation? More specifically, within the region of phase space where $\D < \SI{1000}{m^2/s}$ in \cref{fig:ASCOT} (that mostly coincides with the re-healed flux surface region), which constant value of $\D$ is sufficient to yield negligible \RE current? As seen in \cref{fig:IrLogInverse}, full \RE beam prevention is only achieved somewhere in the range $\D = \SI{10\mydash18}{m^2/s}$. Therefore, compared to the highly diffusive edge region ($\D \approx \SI{10^3\mydash10^5}{m^2/s}$), a relatively small amount of core transport is needed. Importantly, note that the advection coefficient $\A[{\rm m/s}]$ is set to the same value as $D[{\rm m^2/s}]$ in these phase space regions, but almost identical results are found when setting $\A=0$, as diffusion dominates in the narrow radial region of re-healed flux surfaces (as long as $\A[{\rm m/s}]\sim \D[{\rm m^2/s}]$). 
    
    \section{Discussion and summary}\label{sec:discussion}

    From the previous sections, it is clear that zeroing the transport in the core ($r/a < 0.3$) in \DREAM is too conservative and pessimistic, resulting in a $\mysim\SI{1}{MA}$ \RE beam. Even so, it is important to note that this current is 5-6 times less than that expected for an unmitigated \RE beam, i.e. no \REMC, so even this conservative base case could be considered successful. \ASCOT simulations evaluate diffusion coefficients spanning $\D \approx \SI{1\mydash1000}{m^2/s}$ in the core, but encouragingly only $\D \sim \SI{18}{m^2/s}$ is needed in that region to completely suppress a \RE beam. \ppcfadd{Perhaps this is one reason why many tokamaks struggle to generate \RE plateaus via ``natural'' disruptions (as in Alcator C-Mod \cite{Granetz2010,Izzo2011}, for example), and instead resort to special ``recipes,'' although lower plasma current and thus lower avalanching certainly also play a role.} 
    
    However, it is not yet known whether this level of transport is achievable \ppcfadd{in \SPARC}. In \cite{Izzo2022}, it was noted that the degree of field line stochastization predicted by \NIMROD could be affected by several approximations, most notably the presence of a close, ideal wall which tends to \emph{limit} \MHD mode growth. This approximation will be explored further in future resistive-wall studies, and perhaps this minimum $\D$-value will even decrease.

    \ppcfadd{
        We can also approach this from another direction: In what ways can the \REMC design be modified to achieve $\D > \SI{18}{m^2/s}$ throughout the plasma? Perhaps most straightforward, the coil could be moved closer to the plasma and farther from the \VV. This would (i)~improve the plasma-coil mutual coupling and reduce the coil's self-inductance, thereby increasing the induced coil current, (ii)~decrease image currents in the conducting wall, and (iii)~enhance the magnetic perturbation amplitude $\db$ in the core. Perhaps a design metric could be the expected diffusion coefficient computed from vacuum fields \ala \cite{Rechester-Rosenbluth_PRL1978}, $\D \propto (\db)^2$. The coil resistance could also be lowered by changing the coil cross-section, length, and material (resistivity). That said, many other factors constrain the design, like available space, forces and stresses, heating, and more.
    }
    
    As discussed in \cite{Tinguely2021}, both advection and diffusion tend to increase with \RE energy, but there is a roll-over when the energetic electron drift orbits effectively average over large regions of stochasticity (see Figure~3 in \cite{Tinguely2021} or \cite{Carbajal2020,Sarkimaki2022} and others for further details). However, for \REs within healed flux surfaces, perhaps large orbits could lead to ``excursions'' into stochastic fields, thus enhancing transport. For example, KORC simulations in \cite{Carbajal2020} found that \REs with Larmor radii similar to island widths could escape them. 
    In addition, the same electric field accelerating \REs causes them to drift radially \cite{Guan2010}, and this was not accounted for in these \ASCOT simulations, but will be pursued in the future.%
        \footnote{See \cite{Sarkimaki2022} for simulations of passing and trapped \REs during an ITER \CQ, including the effects of collisions and the electric field.}
    
    %\red{Konsta: Are collisions included in your \ASCOT simulations? i.e. energy loss and pitch angle scattering? Maybe the effect would be small.}
    
\begin{figure}[h!]
        \centering
        \begin{subfigure}{\halfwidth}
            \includegraphics[width=\textwidth]{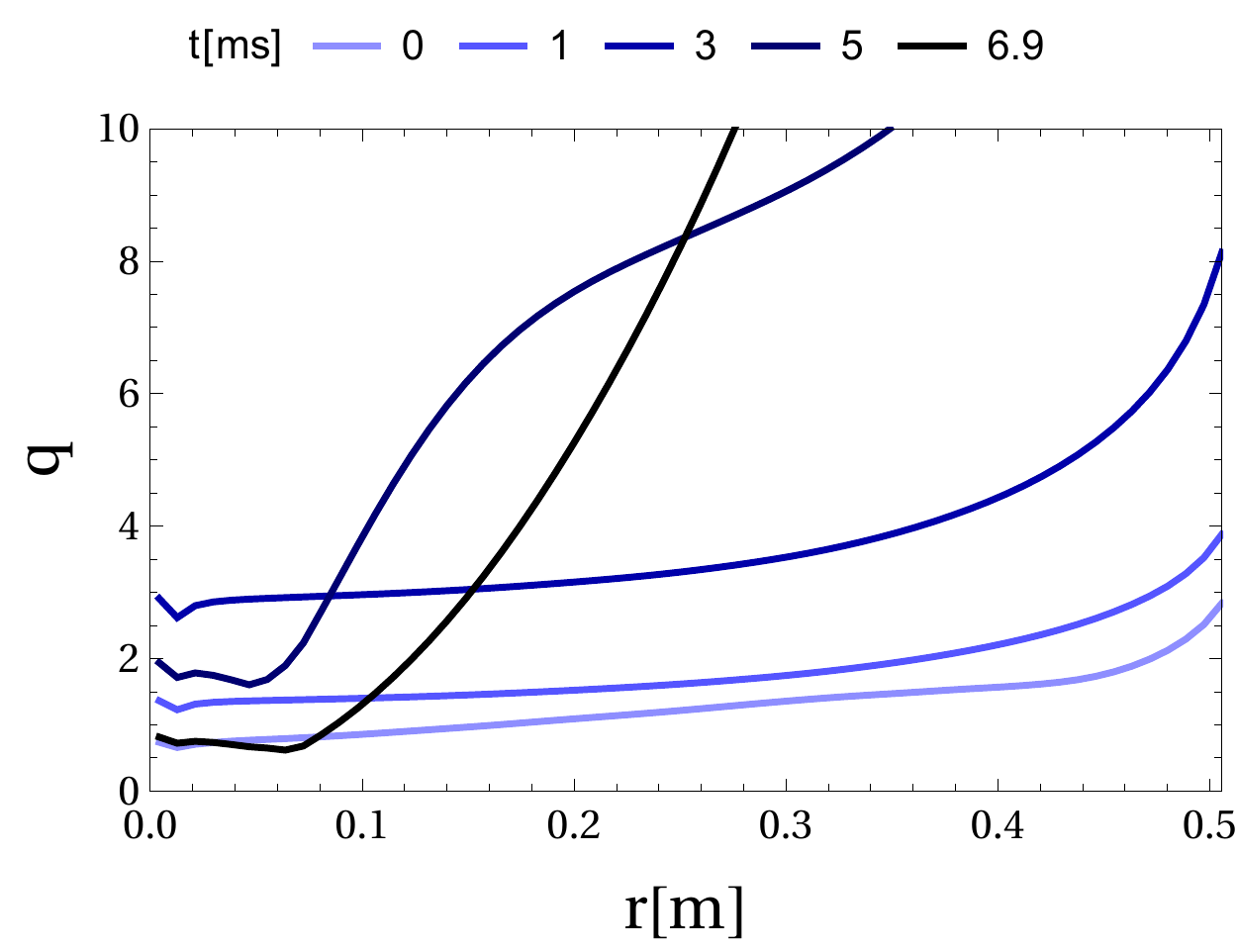}%{figs/qFig.pdf}
            \caption{}
            \label{fig:q}
        \end{subfigure}
        \begin{subfigure}{\halfwidth}
            \includegraphics[width=\textwidth]{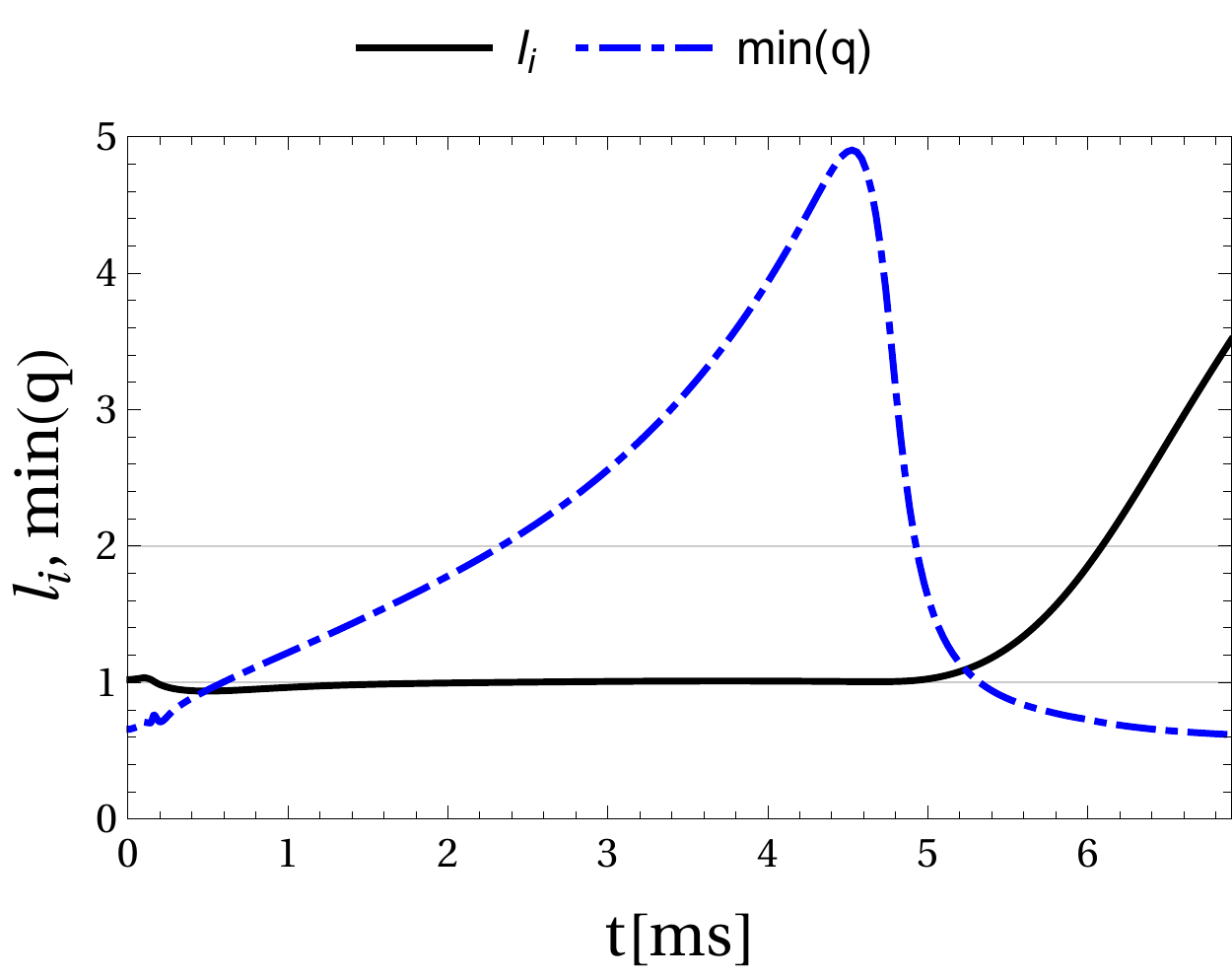}%{figs/liqFig.pdf}
            \caption{}
            \label{fig:li_minq}
        \end{subfigure}
        \caption{\DREAM results for the base case in \cref{fig:IrLog} with $\D=0$ below $\D<\SI{1000}{m^2/s}$: (a)~evolution of the safety factor $\q$-profile vs minor radius for five times, and (b)~time evolution of the minimum $\q$-value (dot-dashed) and internal inductance $\li$ (solid).}
        \label{fig:q_and_li}
    \end{figure}

    Perhaps most importantly, the effect of the \RE population itself on the magnetic field and \MHD has not yet been fully assessed. \Cref{fig:q_and_li} shows the time-evolution of the $\q$-profile, its minimum value, and the internal inductance ($\li$) in \DREAM for the base case. Although slightly later in time than in \cref{fig:NIMROD}, the central safety factor $\qo$ also surpasses $\q=2$; however, unlike the \NIMROD results, the increasing \RE current then reduces $\qo<2$ at $\t \approx \SI{5}{ms}$ and $\qo<1$ at $\t \approx \SI{5.5}{ms}$. Thus, in theory, the \REMC should regain resonance in the core beyond $\t > \SI{5}{ms}$, thereby enhancing transport and reducing the \RE current, but this was not captured in the current workflow. A destructive kink instability might also be expected, as seen in experiment \cite{Cai_2015, PazSoldan2019}, for such low $\qo$ and high $\li$. 
    
   %\red{Val/Konsta/Istvan: There has been some work to include this in NIMROD and other codes, like JOREK, right? Is there anything we can say about this? Or papers we can cite? (Izzo: see above addition)}

    Even then it is not clear what overall effect this self-regulation would have on the \RE beam which already has a current $\mysim\SI{1}{MA}$ by $\t \approx \SI{5}{ms}$ (for the base case with no island transport). Luckily, the \ohmic current has almost completely decayed by then, and the relatively long L/R time (${>}\SI{10}{ms}$) of the \REMC will maintain the coil current and its perturbative effect. Furthermore, any additional transport within the re-healed flux surfaces will help lower this quasi-stationary \RE current. A fluid RE model that could capture this effect has been incorporated into the JOREK \cite{bandaru2019simulating,bandaru2021magnetohydrodynamic} and M3D-C1 \cite{zhao2020simulation} \MHD codes; a similar model is being implemented in \NIMROD \cite{sainterme2020development} and benchmarked against the existing codes. Its application to the \SPARC \REMC will be pursued in future work.

    \section*{Acknowledgments}
    
    Supported by Commonwealth Fusion Systems, Swedish Research Council (Dnr. 2018-03911), US DOE Award Numbers DE-FC02-04ER54698 and DE-FG02-95ER54309. This work has been carried out within the framework of the EUROfusion Consortium, funded by the European Union via the Euratom Research and Training Programme (Grant Agreement No 101052200 -- EUROfusion). Views and opinions expressed are however those of the author(s) only and do not necessarily reflect those of the European Union or the European Commission. Neither the European Union nor the European Commission can be held responsible for them.

   \section*{References}
        \bibliographystyle{unsrt}
        %\bibliography{bib}

\end{document}